# Meteor Beliefs Project: Meteors in the Māori Astronomical Traditions of New Zealand

Tui R. Britton[1] and Duane W. Hamacher[2]

[1]Department of Physics & Astronomy, Macquarie University, Sydney, NSW, 2109, Australia
Email: tui.britton@mq.edu.au

[2]Nura Gili Centre for Indigenous Programs, University of New South Wales, Sydney, NSW, 2052, Australia
Email: d.hamacher@unsw.edu.au

**Abstract**

We review the literature for perceptions of meteors in the Māori cultures of New Zealand. We examine representations of meteors in religion, story, and ceremony. We find that meteors are sometimes personified as gods or children, or are seen as omens of death and destruction. The stories we found highlight the broad perception of meteors found throughout the Māori culture and demonstrate that some early scholars conflated the terms *comet* and *meteor*.

*"The sparks of Rongomai never fail to bring shouts of wonder from the lips of men."*

A description of a Maori ancestor taking the form of a bright meteor (Kingsley-Smith, 1967).

**Introduction**

The Māori of New Zealand (Aotearoa) are a Polynesian people who descend from the Cook Island Māori and other Eastern Polynesian groups (King, 2003). The Māori migrated to Aotearoa in the 13[th] century, travelling by waka (canoe) from Rarotonga (Anderson, 2009; Walter & Moeka'a, 2000). This Great Migration comprised seven or eight wakas[1] containing the ancestors of the present day Māori people of Aotearoa (Evans, 2009). Many Māori can trace their lineage back to one or more of the original waka.

The arrival of Europeans (Pakeha[2]) in the late 18[th] century marked a change in Māori culture. The Māori quickly learnt to read and write, being introduced to these new skills from the Pakeha. The Māori embraced these skills in order to preserve their knowledge and oral traditions. Countless documents exist written by both Māori and Pakeha that contain the knowledge and traditions of the Māori. Researchers are still tracking down and sifting through these documents for many aspects relating to Māori culture, particularly astronomy (e.g. Harris & Matamua, 2012; Orchiston, 2000).

---

[1]There is some contention over the exact number of waka and the island(s) from which they set out on their journey.
[2]Pakeha is not a derogatory term and is used colloquially to refer to anyone of European decent.

We present here a brief examination of the cultural knowledge of Māori astronomy by focusing on their folklore and legends of meteors or shooting stars. We examine many well-known records for references to meteors in story, religion, and ceremony. The majority of published information about Maori astronomical traditions, including meteor names, comes from the work of Elsdon Best (1955). His published study of Maori astronomy remains the most detailed and comprehensive to date.

**Meteor Names**

In Aotearoa, meteors have many names, varying from region to region. In the Bay of Plenty, meteors are known as *matakōkiri* (the darting ones), *tūmatakōkiri*, *kōtiri*, *kōtiritiri*, *tamarau*, and possibly *unahi o Taero* (Stowell, 1911:199; Best, 1955:69). The Ngatiawa tribe near Whakatane say that *Taneatua* - the tohunga (priest) of the Mataatua (one of the seven original waka) - brought comets and meteors with him on the Great Migration and released them into the southern skies (Kingsley-Smith, 1967). The Ngatiawa call these meteors Rongomai, a name also used to denote a comet – in particular Halley's Comet (Stowell, 1911: 200).

**Perception of Meteors**

Perceptions of meteors were diverse among the Māori. Meteors were generally seen as omens of evil or death (Best 1955; Mackrell, 1985:21-28). A meteor may portend the death, or the rise and fall, of a chief (Best, 1955:70). They were also viewed as star children or personifications of supernatural beings or ancestors (*ibid*).

The physical characteristics of a meteor, such as the brightness and trajectory, have special meaning to the Māori (*ibid*). Bright meteors denote good omens, while fainter ones denote evil omens. If a meteor is seen heading toward the observer, it is a good sign (*ibid*). For example, the matakōkiri are stars that have wandered out of their places and have been struck by their elders - the sun and moon. If a matakōkiri appeared to approach a person directly, it was seen as a good omen (*ibid*).

Meteors are sometimes refereed to as Raririki (little shining ones). The twinkling stars are children playing across the robe of Rangi (the sky father). Occasionally one of the children will trip and fall, flashing across the sky in a brilliant light (Reed, 1950:190).

**Stories of Gods**

Meteors are personified as *atua* (supernatural beings; Best, 1955:70). When an atua is expelled from the sky for behaving badly, he is seen as a meteor. Atuas are also known to occasionally visit the earth (Orbell, 1996:165), suggesting a link between meteors and meteorites.

Rongomai was an atua who provided guidance and protection in war. We know that Rongomai is used to refer to Halley's Comet (Tregear 1891:425). But Rongomai was also known to "move through space" and "give off sparks". Best (1955:67) cites an account by Rev. R. Taylor who claimed that when the Pakakutu pa (fort) at Otaki was besieged, Rongomai was seen in broad daylight as a *"fiery form rushing through space"* striking the ground and causing dust to rise. This description clearly illustrates a fireball and subsequent meteorite impact and not a comet. Otaki is approximately 65 km north of Wellington on the western coast of the North Island. Best also describes a place named *Te Hapua o Rongomai* at Owhiro Bay, south of Wellington, where an atua is said to have descended to earth. The motif of a star falling from the sky and impacting in the distance during or preceeding a war or battle is found across the world (e.g. Avilin, 2007). This may not actually describe a witnessed event, but may simply be an idea incorporated into folklore.

Tutaka, one of Best's male informants from the Tuhoe tribe, stated that Tunui is not a star but a demon - a spirit that flies through space and has a "big head" (Best, 1955:68). Best categorizes Tunui as a comet, but the description clearly indicates that Tunui is a bright meteor or fireball. The appearance of Tunui signals that someone has died. This is also a common perception among Aboriginal groups in northern Australia (Hamacher & Norris, 2010).

In many early writings about astronomical traditions, comets and meteors are often conflated (see Hamacher & Norris, 2010). Another story recorded by Best (1955:68) highlights this. He describes Tunui and Te Po-tuatini as spirits that fly through space, which Best identifies as comets. They are seen in the night sky and they are atua toro - inquisitive, reconnoitring gods. Their human mediums (usually the tohunga) placate and influence them by means of a ritual saying. Thus, Tunui is employed as a war-god and certain invocations are addressed to him. Comets do not appear to "fly" through space, nor are they fleeting. Instead, they gradually move across the sky from night to night. It is clear that what is described is a meteor and not a comet.

**Stories of Ancestors & Men**

An ancestor and supernatural being named Tūmatakōkiri is seen as a meteor, according to an "old warlock" of the sons of Awa (Best, 1955:70). Tūmatakōkiri foresees the positions of celestial bodies, and the seasonal and weather conditions, as he flies through the skies (*ibid*). If he moves downward, the following season will be a windy. If he maintains a level trajectory, the following season will be successful and bear much fruit.

Hape from Ohiwa is the ancestor of the Te Hapu Oneone people ("*the people of the soil*", Orbell, 1996:45). He had two sons, Tamarau and Rawaho. Rawaho was the eldest and a tohunga, however, it was Tamarau that entered the house of his fathers' death first and inherited his father's powers. This turned him into an atua and gave him the power of flight; he would fly around from place to place. Tamarau lived in Kawekawe but if anyone approached his house he would turn into a meteor and fly away.

In Maori astronomical traditions, the bright star Sirius (Alpha Canis Major) is called Rehua (Stowell, 1911:201-202). The star is said to have come as a *"flaming star from out of the dark-hole"*, a reference to the Coalsack nebula near the Southern Cross. Rehua flew across the sky with *"lighting speed"*, venturing among the stars before finally settling in his current place in the sky. The motif of a flaming star emerging from the Coalsack is also found in Aboriginal traditions of Australia (Hamacher & Norris, 2010).

**Stories of Destruction**

Māori mythology is rife with connections between fire, the disappearance of the moa (a large flightless bird akin to an ostrich or emu and is now extinct), and an object falling from the sky (Snow, 1983; Steel & Snow, 1992; Bryant, 2001). Meteors were believed to bring fire to the earth, suggesting a cultural memory of an airburst or meteorite impact event. According to Steel & Snow, the word moa itself is recent; in an early period – *before the flames* – the Māori called the bird *Pouakai*, but later it was called *Manu Whakatau*. One translation of this is *bird felled by strange fire*. The following Māori poem highlights this view (Steel & Snow, 1992:571):

> *"Very calm and placid have become the raging billows*
> *That caused the total destruction of the moa*
> *When the horns of the Moon fell from above down"*.

Steel & Snow (*ibid*) cite a report of a conversation with an 88-year-old Māori chief who claimed that:

> *"The moa disappeared after the coming of Tamaatea (a man/god) who set fire to the land. The fire was not the same as our fire but embers sent by Rangi (the sky). The signs of the fires are still to be seen where red rocks like berries are found."*

Attempts to directly relate these oral traditions to a meteorite impact have been problematic and contentious. In 2003, Dallas Abbott and her colleagues reported the discovery of a putative submarine impact crater 250 km south of Stewart Island (48.3 S, 166.4 E) with a diameter of 20±2 km that they believe impacted in 1443 AD (Abbott et al, 2003, 2005). Abbott and her colleagues named the structure Mahuika, after the Māori god of fire, believing this to be the impactor that sparked the Māori traditions described in this section.

Impacts large enough to create a crater 20 km wide are believed to impact the Earth once every three million years (Collins et al, 2005), casting doubt on the proposed date of 1443 CE. The impact hypothesis relating to the Mahuika structure and the Māori traditions has been challenged (Goff et al. 2003; 2010) but remains a topic of contentious debate (Bryant et al. 2007). An impact origin of the structure is still awaiting confirmation.

**Discussion**

Some of the stories in Māori traditions seem to describe a meteorite fall or impact. Only nine meteorite finds have been confirmed in New Zealand. In order of discovery, they are Wairarapa Valley (1863: Find), Makarewa (1879: Find), Mokoia (1908: Observed Fall), Morven (1925: Find), View Hill (1953: Find), Waingaromia (1970: Find), Duganville (1976: Find), Kimbolton (1976: Find) (Grady, 2000), and Ellerslie (2004: Observed Fall). There is currently no confirmed connection between known meteoritic events and those recorded in Māori traditions. Meteor traditions that may describe an impact from Otaki and Owhiro Bay are in the same general region as the Wairarapa Valley meteorite-find in 1863, but any connection between them is speculative. There is no confirmed impact crater associated with the proposed impact event that describes the destruction of the moa. There are no known reports of the Māori using meteoritic material for practical or social purposes. However, these are topics of current research.

**Conclusion**

We have highlighted various views of meteors in Māori stories, religion, and ceremony. The Māori have a broad perception of meteors. We find that they often view meteors as atua (supernatural beings) or ra ririki (children of light). However, meteors are also synonymous with fire and destruction. This is similar to other cultures around the world where meteors are viewed as bad omens (e.g. Hamacher & Norris, 2010). We also find that references to comets and meteors are often conflated, with descriptions of meteors being mistaken for comets.

**Acknowledgements**

We acknowledge the Māori of Aotearoa, both past and present and wish to thank Alastair MacBeath. T. Britton is supported by the Macquarie University Research Excellence Scholarship.

**About the Authors**

Tui Britton is finishing a PhD in astrophysics at Macquarie University in Sydney, Australia on the topic of massive star formation. Born in New Zealand and raised in Singapore, she graduated with an International Baccalaureate from the United World College of Southeast Asia and earned BSc and MSc degrees in astrophysics from Michigan State University and the University of New South Wales, respectively. She also works as an astronomy educator at Sydney Observatory. Tui is of Maori decent on her maternal side and conducts research on Maori and Polynesian astronomy.

Dr. Duane Hamacher is a Lecturer in the Nura Gili Indigenous Centre at the University of New South Wales in Sydney, Australia. He publishes extensively on cultural and historical aspects of astronomy and meteoritics, with a focus on Indigenous Australia. Born in the United States, he earned BSc and MSc degrees in astrophysics from the University of Missouri and the University of New South Wales, respectively, before completing a PhD at Macquarie University studying Aboriginal astronomy. He also works as a consultant curator and astronomy educator at Sydney Observatory.